\documentclass[reqno,11pt]{amsart}
\usepackage{hyperref}
\usepackage[mathscr]{eucal}
\usepackage{enumerate}
\numberwithin{equation}{section}

\newtheorem{Def}{Definition}[section]
\newtheorem{Thm}[Def]{Theorem}

\newtheorem{Lemma}[Def]{Lemma}

\newcommand{\beq}{\begin{equation}}
\newcommand{\eeq}{\end{equation}}
\newcommand{\Proof}{\begin{proof}}
\newcommand{\QED}{\end{proof} \noindent}

\newcommand{\R}{\mathbb{R}}

\newcommand{\nn}{\nonumber}

\allowdisplaybreaks[1]

\title[Relativistic Damped Euler Equations]{A Proposal of a Damping Term for the Relativistic Euler Equations} 

\author[M.\ Reintjes]{Moritz Reintjes}
\address{IMPA - Instituto Nacional de Matem{\'a}tica Pura e Aplicada \\ Rio de Janeiro, 22460-320, Brasil}
\email{moritzreintjes@gmail.com}
\thanks{M. Reintjes is a Post-Doctorate at IMPA, funded through CAPES-Brazil.}

\begin{document}

\begin{abstract}
We introduce a damping term for the special relativistic Euler equations in $3$-D and show that the equations reduce to the non-relativistic damped Euler equations in the Newtonian limit ($c\rightarrow \infty$). We then write the equations as a symmetric hyperbolic system for which local-in-time existence of smooth solutions can be shown.  
\end{abstract}

\maketitle


\section{Introduction}

The non-relativistic damped Euler equations are given by \cite{Sideris}
\begin{eqnarray}
\rho_t + \nabla \cdot(\rho v) &=& 0 \label{Euler_cl_mass} \\
\rho \left( v_t + v\cdot \nabla v \right) + \nabla p &=& - a \rho v \label{Euler_cl_mom} , 
\end{eqnarray}
where $\nabla=(\partial_x,\partial_y,\partial_z)$ denotes the gradient in Cartesian coordinates on $\R^3$, $a$ is a positive constant, $\rho$ is the mass-density, $v$ is the fluid velocity and $p$ is the pressure, which is assumed to be a given function of $\rho$. The damping term is given by $a \rho v$.

The system \eqref{Euler_cl_mass} - \eqref{Euler_cl_mom} models flow of fluids or gases through some fixed background material which slows down the fluid flow (for positive $a$). For example, flow of a fluid through soil or flow of a light fluid or gas through a heavier fluid, for instance air bubbles moving through water. Further examples, with fluid velocities on the order of the speed of light, could be radioactive radiation emitted by the sun passing through the atmosphere of the earth or neutrino radiation passing through stellar matter during gravitational collapse triggering a supernovae. (See \cite{BieriGarfinkle} for a fluid model of neutrino radiation.) However, for such large velocities, the description by \eqref{Euler_cl_mass} - \eqref{Euler_cl_mom} is insufficient as relativistic effects become increasingly dominant. The objective of this paper is to derive a damping term for the relativistic Euler equations which reduces to the one in \eqref{Euler_cl_mom} in the non-relativistic limit. 

In Section \ref{Sec: Euler_mass-energy} we propose a relativistic damping term proportional to mass-energy density. The frame splitting of the resulting damped relativistic Euler equations is computed in Section \ref{Sec: Splitting_1}, which is the starting point for computing their Newtonian limit in Section \ref{Sec: Newtonian-limit_1}. To prove their local well-posedness with Kato's method \cite{Kato}, one needs to write the damped relativistic Euler equations as a symmetric hyperbolic system, which is accomplished in Section \ref{Sec: symmetrization}. In Section \ref{Sec: Euler_particle-number} we discuss a damping term proportional to the particle-number density and compute the Newtonian limit of the resulting equations, from which we conclude that such a damping seems unphysical.

\section{The Special Relativistic Damped Euler Equations} \label{Sec: Euler_mass-energy}

We propose the Relativistic Damped Euler equations to be given by
\beq \label{Euler_rel-damping}
\text{div}\, T = K, 
\eeq
for
\beq \label{damping-4-force}
K^\mu =  -a \, \gamma(v)  \left( \begin{array}{c} \frac{1}{c} \vec{v}^2 \cr \vec{v} \end{array} \right)\epsilon\, ,
\eeq
where $c$ denotes the speed of light, $\epsilon$ is the (relativistic) mass-energy-density of the fluid, $K^\mu$ is the Lorentz-force of the classical damping term in \eqref{Euler_cl_mom} and 
\beq \nn
\gamma(v)^{-1} \equiv \sqrt{1-\frac{v^2}{c^2}}.
\eeq  
Here $T$ is the energy-momentum tensor of a perfect fluid,         
\beq \label{T_perfect_fluid}
T^{\mu\nu} = \left( \epsilon + p \right) u^\mu u^\nu + p \eta^{\mu\nu}, 
\eeq
where $p$ denotes the pressure and $u^\mu$ the fluid four-velocity normalized to 
\beq \label{normalization}
u^\sigma u_\sigma = -1.
\eeq 
The divergence is taken with respect to coordinates $(x^0,...,x^3)$ and we raise and lower indices with the Minkowski metric $\eta_{\mu\nu}$, given by $\eta_{00}=-1$ and $\eta_{ij} = \delta_{ij}$ for the Kronecker delta $\delta_{ij}$ for $i,j \in \{ 1,2,3\}$, c.f. \eqref{Minkowski-metric}. 
Note, as shown in \eqref{Prel_relation_u_v_3D}, $\vec{v}$ is given in terms of $u^\mu$ through  
\beq \label{relation_u_v_3D}
\vec{v}^i = c\frac{u^i}{u^0} .
\eeq

A peculiarity of \eqref{Euler_rel-damping} is that $K$ is proportional to mass-energy-density $\epsilon$, however, by the usual connection between Lorentz force and classical force, one would naively expect the classical mass-density $\rho$ to enter but not $\epsilon$. The reason why $\rho$ cannot appear in \eqref{Euler_rel-damping} is that mass is equivalent to energy in Relativity, so that considering a mass-density alone does not make sense. One might be tempted at this point to introduce the particle number as an additional fluid variable (and augment the above equations by its conservation law), since the particle number density can be interpreted as rest mass density of the fluid. However, as shown in Section \ref{Sec: Euler_particle-number}, one does not recover \eqref{Euler_cl_mom} from the resulting equations in the non-relativistic limit. Moreover, the naive choice of $a\epsilon u^\mu$ as a relativistic damping term would result in a damping in the conservation of mass equation and not in the balance of momentum equation. We therefore propose \eqref{Euler_rel-damping} as the Relativistic Damped Euler equations. (Let us remark, that it also seems reasonable to allow for $\epsilon$ to enter \eqref{damping-4-force} non-linearly as $\epsilon^\alpha$ for some $\alpha>0$, however, we only focus on the linear case here.) We wonder whether this type of damping, based on a Minkowski force proportional to $\epsilon$ or $\epsilon^\alpha$, is indeed unique.

\section{Their Frame Splitting}\label{Sec: Splitting_1}

We now compute the components of \eqref{Euler_rel-damping} along $u^\mu$ and orthogonal to $u^\mu$. To begin, a straightforward computation yields that \eqref{Euler_rel-damping} is equivalent to
\beq \label{Euler_mass-energy-damping_again}
\left( \epsilon + p \right) u^\mu u_{\nu ,\mu} + u_\nu \left( (\epsilon + p) u^\mu  \right)_{,\mu} + p_{,\nu} = K_\nu ,
\eeq
where we use a comma to denote differentiation, e.g., $u_{\nu ,\mu}\equiv \partial_\mu u_\nu$. Before we contract with $u^\mu$, let us remark that $u^\mu$ and $\vec{v}$ are related by 
\beq \label{relation_u_v}
u^\mu  = \frac{\gamma(v)}{c}\left( \begin{array}{c} c \cr \vec{v}  \end{array} \right) \equiv \frac{\gamma(v)}{c} \ v^\mu,
\eeq 
c.f. \eqref{Prel_relation_u_v}, which together with \eqref{damping-4-force} implies
\beq \nonumber
K^\sigma u_\sigma = 0  .
\eeq
Moreover, observe that \eqref{normalization} implies
\beq \nn
 u_\mu \, u^\mu_{\ ,\nu}= 0.
\eeq 
Now, contracting \eqref{Euler_mass-energy-damping_again} with $u^\nu$, we obtain
\beq \label{Euler_mass-energy-damping_splitting_1}
\epsilon_{,\sigma} u^\sigma  + \left(  \epsilon + {p} \right) u^\sigma_{\ ,\sigma} = 0 .
\eeq
This is the relativistic balance of mass-energy equation.

To continue, we introduce the orthogonal projection  
\beq \nonumber
\Pi^{\mu\nu} \equiv \eta^{\mu\nu} + u^\mu u^\nu ,
\eeq
for which a straightforward computation gives us  
\begin{eqnarray} \nn
\Pi_{\mu\nu} u^\nu &=& 0, \cr
\Pi_{\mu\nu} u^\nu_{\ ,\sigma} u^\sigma &=& u_{\mu,\sigma} u^\sigma .
\end{eqnarray} 
Now, contracting \eqref{Euler_mass-energy-damping_again} with $\Pi_{\mu\nu}$ and using the previous two identities yields
\beq \label{Euler_mass-energy-damping_splitting_2}
\left(  \epsilon + p \right) u^\mu_{\ ,\sigma} u^\sigma + \Pi^{\mu\nu} p_{,\nu} =  \Pi^{\mu\nu} K_\nu .
\eeq
This is the relativistic balance of momentum equation. To summarize, the damped Euler equations \eqref{Euler_rel-damping} are equivalent to \eqref{Euler_mass-energy-damping_splitting_1} and \eqref{Euler_mass-energy-damping_splitting_2}.

\section{Their Newtonian Limit}\label{Sec: Newtonian-limit_1}

We now take the Newtonian limit, $c\rightarrow \infty$, of \eqref{Euler_mass-energy-damping_splitting_1} and \eqref{Euler_mass-energy-damping_splitting_2} in a formal sense, and show that the equations approach the classical damped Euler equations, \eqref{Euler_cl_mass} - \eqref{Euler_cl_mom}.

To begin, we derive some useful relations. A direct computation shows that
\beq\nonumber
\partial_\sigma \gamma(v) = \frac1{c^2} \gamma(v)^3 \vec{v} \cdot \partial_\sigma \vec{v} = O(c^{-2}) .
\eeq
Using \eqref{relation_u_v} and that $\partial_0 = c^{-1}\partial_t$, (which follows from $x^0 \equiv ct$), we find that
\beq \label{identity_1}
u^\sigma_{\ , \sigma} = \frac{\gamma(v)}{c} \nabla \cdot \vec{v} + O(c^{-2}).
\eeq

To continue, note that $\epsilon$ in \eqref{T_perfect_fluid} is assumed to be given in units of energy and can be replaced by a mass-energy density $\rho$ in units of mass, by identifying $\epsilon \equiv \rho c^2$. We obtain
\begin{eqnarray}\nonumber
\epsilon_{,\sigma} u^\sigma 
&=& \frac{\gamma(v)}{c} \left( c\, \partial_{0} \epsilon + \vec{v} \cdot \nabla \epsilon  \right) \cr
&=& \frac{\gamma(v)}{c} \left( \partial_{t} \epsilon + \vec{v} \cdot \nabla \epsilon  \right)  \cr
&=& c\gamma(v) \left( \partial_{t} \rho + \vec{v} \cdot \nabla \rho  \right) 
\end{eqnarray}
Substituting the above identities into \eqref{Euler_mass-energy-damping_splitting_1}, we obtain
\beq \nonumber
c\gamma(v) \left( \partial_{t} \rho + \vec{v} \cdot \nabla \rho  \right)
  + \left( \rho c^2 + p \right) \left( \frac{\gamma(v)}{c} \nabla \cdot \vec{v} + O(c^{-2}) \right) = 0 .
\eeq
Dividing by $c$, taking the limit $c\rightarrow \infty$ and using that $\gamma(v) \rightarrow 1$ as $c\rightarrow \infty$, the above equation reduces to
\beq \label{Euler_mass-energy-damping_splitting_1_limit}
 \partial_{t} \rho + \vec{v} \cdot \nabla \rho  +  \rho   \nabla \cdot \vec{v}  = 0 .
\eeq
This is equivalent to the conservation of mass equation of the non-relativistic Euler equations, \eqref{Euler_cl_mass}, and allows us to interpret $\rho$ as (classical) mass density.

We now take the limit $c\rightarrow \infty$ of \eqref{Euler_mass-energy-damping_splitting_2}. For this, observe that by \eqref{relation_u_v},                   
\begin{eqnarray}       
u^\nu_{\ ,\sigma} u^\sigma 
&=& \frac{\gamma(v)^2}{c^2} v^\sigma \partial_\sigma v^\nu  +  \frac{\gamma(v)}{c^2} v^\nu v^\sigma \partial_\sigma \gamma(v)  \cr
&=& \frac{\gamma(v)^2}{c^2} v^\sigma \partial_\sigma v^\nu \, +\, O(c^{-3})  \cr
&=& \frac{\gamma(v)^2}{c^2} \left( \begin{array}{c} 0 \cr  \partial_t \vec{v} + \vec{v} \cdot \nabla \vec{v} \end{array} \right) + O(c^{-3})  .
\end{eqnarray}
Substituting the above identity into \eqref{Euler_mass-energy-damping_splitting_2}, we obtain
\beq \label{Euler_mass-energy-damping_splitting_2_limit_aux}
\left(  \rho + \frac{p}{c^2} \right)  \left( \gamma(v)^2 \left( \begin{array}{c} 0 \cr  \partial_t \vec{v} + \vec{v} \cdot \nabla \vec{v} \end{array} \right) + O(c^{-1})  \right) 
+ \Pi^{\mu\nu} p_{,\nu} = \Pi^{\mu\nu} K_\nu .
\eeq
A straightforward computation shows that
\beq\lim_{c\rightarrow \infty} \Pi_{\mu\nu} = \left( \begin{array}{cc} 0 & 0 \cr 0 & \text{id}_3  \end{array}  \right),
\eeq 
where $\text{id}_3$ denotes the identity on $\R^3$. Thus, taking the limit of \eqref{Euler_mass-energy-damping_splitting_2_limit_aux} yields
\beq \label{Euler_mass-energy-damping_splitting_2_limit}
 \rho  \partial_t \vec{v} + \rho \vec{v} \cdot \nabla \vec{v}  + \nabla p = \vec{K} ,
\eeq
where $\vec{K} = - a \epsilon \vec{v}$. This is the non-relativistic balance of momentum equation \eqref{Euler_cl_mom}.

\section{Symmetrization and Local Existence} \label{Sec: symmetrization}

In this section we write the Euler equations as a symmetric hyperbolic system, c.f. \cite{BrauerKarp}. Subsequently, we work in units where $c=1$ and we assume an equation of state $p= A \epsilon^\gamma$ for $A>0$ and $\gamma> 1$. Moreover, suppose that $\epsilon >0$ and that $\sigma \equiv \sqrt{p^\prime(\epsilon)}\leq1$. Recall the Euler equations in their frame splitting, \eqref{Euler_mass-energy-damping_splitting_1} and \eqref{Euler_mass-energy-damping_splitting_2}, 
\begin{eqnarray} \label{Euler1}
\epsilon_{,\sigma} u^\sigma  + \left(  \epsilon + p \right) u^\sigma_{\ ,\sigma} &=& 0,  \cr 
\left(  \epsilon + p \right) u^\mu_{\ ,\sigma} u^\sigma + \Pi^{\mu\nu} p_{,\nu} &=&  \Pi^{\mu\nu} K_\nu .
\end{eqnarray}
To begin symmetrizing \eqref{Euler1}, observe that by $u_\mu {u^\mu}_{,\sigma} =0$ the following relations holds
\begin{eqnarray} \nn
{u^\mu}_{,\sigma} u^\sigma  &=&  \Pi^{\mu\nu} u^\sigma u_{\nu,\sigma} 
= \left( \Pi^{\mu\nu}  + u^\mu u^\nu  \right) u^\sigma u_{\nu,\sigma}, \cr
u^\sigma_{\ , \sigma} &=& \eta^{\rho\sigma} u_{\rho,\sigma} = \Pi^{\rho\sigma} u_{\rho,\sigma}.
\end{eqnarray}
Substituting the previous relations into \eqref{Euler1}, we write \eqref{Euler1} as
\begin{eqnarray} \label{Euler2}
\epsilon_{,\rho} u^\rho  + \left(  \epsilon + p \right)  \Pi^{\nu\rho} u_{\nu,\rho} &=& 0,  \cr 
\sigma^2 \Pi^{\mu\rho} \epsilon_{,\rho}   +  \left(  \epsilon + p \right) \hat{\Pi}^{\mu\nu} u^\rho u_{\nu ,\rho}  &=&  \Pi^{\mu\nu} K_\nu ,
\end{eqnarray}
where $\sigma \equiv \sqrt{p^\prime(\epsilon)} = \sqrt{A\gamma} \epsilon^{\frac{\gamma-1}{2}}$ and 
\beq\label{velocity-flip}
\hat{\Pi}^{\mu\nu} \equiv {\Pi}^{\mu\nu} + u^\mu u^\nu . 
\eeq
To continue, we introduce the Makino variable \cite{Makino}
\beq\label{Makino-var}
w \equiv \tfrac{2}{\gamma-1}\, \sigma ,
\eeq
from which we find 
\beq \nn
w^\prime(\epsilon) \equiv \frac{dw}{d\epsilon} = \sqrt{A\gamma} \epsilon^{\frac{\gamma-3}{2}}.
\eeq
Setting
\beq\nn
\kappa \equiv \frac{\epsilon}{\epsilon + p} = \frac{1}{1+A\epsilon^{\gamma-1}} = \frac{4\gamma}{4\gamma + (\gamma -1)^2 w^2},
\eeq
multiplying the first equation in \eqref{Euler2} with $\kappa^2 w^\prime (\epsilon)$ and dividing the second equation by $\epsilon+p$, we write \eqref{Euler2} as
\begin{eqnarray} \label{Euler2b}
\kappa^2 w_{,\rho} u^\rho  + \kappa^2 \left(  \epsilon + p \right)w^\prime(\epsilon) \Pi^{\nu\rho} u_{\nu,\rho} &=& 0,  \cr 
\tfrac{\sigma^2}{\epsilon+p} \tfrac{1}{w^\prime(\epsilon)} \Pi^{\mu\rho} w_{,\rho}   +   \hat{\Pi}^{\mu\nu} u^\rho u_{\nu ,\rho}  &=& \tfrac{1}{\epsilon+p} \Pi^{\mu\nu} K_\nu .
\end{eqnarray}
Now, a straightforward computation shows that
\begin{eqnarray}\nn
\kappa^2 (\epsilon+p)w^\prime(\epsilon) 
&=& \frac{\epsilon^2}{\epsilon + p} w^\prime(\epsilon) \cr 
&=& \frac{\sqrt{A\gamma}}{1 + A\epsilon^{\gamma-1}} \epsilon^{\frac{\gamma-1}{2}} \cr
&=&  \kappa \tfrac{\gamma-1}{2} w  \cr
&=& \kappa \sigma 
\end{eqnarray}
and thus 
\begin{eqnarray}\nn
\frac{\sigma^2}{\epsilon+p} \frac{1}{w^\prime(\epsilon)} 
= \kappa \sigma, 
\end{eqnarray}
so that \eqref{Euler2b} simplifies to
\begin{eqnarray} \nn
\kappa^2 u^\rho w_{,\rho}   + \kappa \sigma   \Pi^{\nu\rho} u_{\nu,\rho} &=& 0,  \cr 
\kappa \sigma \Pi^{\mu\rho} w_{,\rho}   +   \hat{\Pi}^{\mu\nu} u^\rho u_{\nu ,\rho}  &=& \tfrac{1}{\epsilon+p} \Pi^{\mu\nu} K_\nu .
\end{eqnarray}
Thus, written in matrix form, \eqref{Euler1} is equivalent to 
\beq \label{Euler3}
A^\rho \ \partial_{\rho} \left( \hspace{-.2cm} \begin{array}{c} w \cr u_\nu   \end{array} \hspace{-.2cm}\right) 
= \frac{1}{\epsilon+p} \left( \hspace{-.2cm}\begin{array}{c} 0 \cr {\Pi^\mu}_{\sigma} K^\sigma   \end{array} \hspace{-.2cm}\right),
\eeq 
for the $5\times 5$ matrices
\beq\label{Euler_symb_symm}
A^\rho =
         \left(\hspace{-.2cm} \begin{array}{cc} \kappa^2 u^\rho 
&  \kappa \sigma \Pi^{\nu\rho} \cr
\kappa \sigma \Pi^{\mu\rho} 
&  \hat{\Pi}^{\mu\nu} u^\rho  \end{array}  \hspace{-.2cm}\right),
\eeq
for $\sigma =0,...,3$. We use $\mu,\nu =0,...,3$ as indices of the matrix coefficients and for the sake of matrix multiplication we consider the components of co-vectors of $\R^{1,3}$ as the lower four components of vectors in $\R^5$.

Obviously the $A^\rho$ in \eqref{Euler_symb_symm} are symmetric matrices. To show that \eqref{Euler3} is a symmetric hyperbolic system, it remains to prove that $A^0$ is positive definite, which is accomplished in the following theorem. 

\begin{Thm} \label{symmetric_hyp_Thm}
Assuming $u^\sigma u_\sigma =-1$, \eqref{Euler3} with \eqref{Euler_symb_symm} is a symmetric hyperbolic system. Moreover, $\epsilon >0$ if and only if $w>0$, and in case that $\epsilon$ and $w$ are positive, then \eqref{Euler1} and \eqref{Euler3} are equivalent. 
\end{Thm}
\Proof
The equivalence of the positivity of $\epsilon$ and $w$ follows from \eqref{Makino-var} and the equivalence of \eqref{Euler1} and \eqref{Euler3} follows from the above computation. 

To prove that \eqref{Euler3} is a symmetric hyperbolic system, we need to show that $A^0$ is positive definite. 
To begin, we prove that $\hat{\Pi}^{\mu\nu}$ is positive definite. Observe that $u_\mu \hat{\Pi}^{\mu\nu} u_\nu = 1$ and $\zeta_\mu \hat{\Pi}^{\mu\nu} \zeta_\nu = \zeta^\mu \zeta_\mu>0$ for any $\zeta$ with $\zeta^\mu u_\mu =0$, which implies
\beq \nn
(a u_\mu + b \zeta_\mu) \hat{\Pi}^{\mu\nu} (a u_\nu + b \zeta_\nu) = a^2 + b^2 \zeta ^\mu \zeta _\mu >0,
\eeq
for $a,b\in \R$, since $\zeta_\mu \hat{\Pi}^{\mu\nu} u_\nu = -\zeta_\mu u^\mu =0$. Since any vector\footnote{For the sake of matrix multiplication with the $5\times 5$ matrix in \eqref{Euler_symb_symm} we here consider the components of co-vectors of $\R^{1,3}$ as the components of vectors in $\R^4$.} $v\in\R^{4}$ can be written as $v_\mu =a u_\mu + b \zeta_\mu$, it follows that $\hat{\Pi}^{\mu\nu}$ is indeed positive definite. 

To continue, we multiply an arbitrary vector $\left(\hspace{-.2cm}\begin{array}{cc} \alpha & v_\mu \end{array} \hspace{-.2cm}\right)\in\R^5$, (for $\alpha \in \R$ and $v_\mu =a u_\mu + b \zeta_\mu$), and its transpose to $A^0$ and compute
\begin{eqnarray} \nn
 \left(\hspace{-.2cm} \begin{array}{cc} \alpha & ^T(v_\mu) \end{array} \hspace{-.2cm}\right) \hspace{-.15cm}
\left(\hspace{-.2cm} \begin{array}{cc} \kappa^2 u^\rho &  \kappa \sigma \Pi^{\mu\rho} \cr
\kappa \sigma \Pi^{\mu\rho} &  \hat{\Pi}^{\mu\nu} u^\rho  \end{array}  \hspace{-.2cm}\right) 
\hspace{-.2cm} \left(\hspace{-.2cm} \begin{array}{c} \alpha \cr v_\mu \end{array} \hspace{-.2cm}\right)  
&=& u^0 \left( \kappa^2 \alpha^2+ v_\mu \hat{\Pi}^{\mu\nu} v_\nu \right)  + 2 \sigma \alpha \kappa \Pi^{\nu 0} v_\nu   \cr
& =& u^0 \left( \kappa^2 \alpha^2+ a^2 + b^2 \zeta^\mu \zeta_\mu \right)  + 2 \sigma \alpha \kappa b \zeta^0.
\end{eqnarray}
Assuming without loss of generality that $u^0$ is positive, only the last term in the previous equation could possibly be negative, however, its absolute value is bounded by the first terms, as we now show: Observe that $\sigma \leq 1$, that $2|\kappa \alpha|\cdot|b \zeta^0| < |\kappa \alpha|^2+|b \zeta^0|^2$ and that $u^0=\sqrt{1+ u^\alpha u_\alpha}>1$, from which we obtain the estimate
\beq \nn
2\sigma |\kappa \alpha|\cdot|b \zeta^0| < \big((\alpha\kappa)^2 + (b\zeta^0)^2 \big) u^0 .
\eeq
Since $\Pi_{\mu\nu}$ is a Riemannian metric on the spacelike hypersurface of vectors orthogonal to $u^\mu$, we finally obtain that $\big(\zeta^0\big)^2 \leq \zeta^\mu \Pi_{\mu\nu} \zeta^\nu = \zeta^\mu \zeta_\mu$. In summary, we conclude that $A^0$ is positive definite and that \eqref{Euler1} is a symmetric hyperbolic system.
\QED

We have shown that \eqref{Euler3} is a symmetric hyperbolic system. Thus, considering \eqref{Euler3} as a $5\times5$ system one could in principle apply Kato's existence theory \cite{Kato} to prove local existence of solutions. However, since the normalization $u^\sigma u_\sigma =-1$ (which is necessary to show that \eqref{Euler3} is symmetric hyperbolic) removes one degree of freedom from the unknowns $w$ and $u$, \eqref{Euler3} appears overdetermined. The resolution here comes from the normalization condition $u^\sigma u_\sigma=-1$ being propagated by \eqref{Euler3} whenever it holds initially, as shown in the following lemma.

\begin{Lemma} \label{normalization_Lemma}
Assume that $ \epsilon + p\neq 0$ and that $u^\sigma u_\sigma=-1$ at some point $p$. The balance of momentum equations in \eqref{Euler2} then implies $u^\sigma \partial_\sigma (u^\nu u_\nu)=0$
at $p$. Thus, $u^\sigma u_\sigma=-1$ holds everywhere along the flow line through $p$.
\end{Lemma}
\Proof
Contracting the second equation in \eqref{Euler2} with $u_\mu$, using $\Pi^{\mu\nu}u_\nu =0$, we find that $u_\mu u^\sigma u^\mu_{\ ,\sigma} =0 $ and this proves the first claim of the lemma.

Setting $f(\tau)\equiv u^\sigma\circ \gamma(\tau) u_\sigma\circ \gamma(\tau)$ for $\gamma$ being a flow line of $u$ through $p$, that is, $\gamma^\prime(\tau)=u\circ\gamma(\tau)$ and $\gamma(0)=p$. Then $f$ satisfies the ordinary differential equation
\beq \nn
\frac{df}{d\tau} = u^\sigma \partial_\sigma (u^\nu u_\nu), \hspace{1cm} f(0)=-1.
\eeq
Since $\frac{df}{d\tau}(0)$ vanishes by the first part of this lemma, we can solve the above ODE by setting $f(\tau)=-1$ for each $\tau$, and since solutions of regular ODE's are unique, we proved the lemma.
\QED

From Theorem \ref{symmetric_hyp_Thm} together with Lemma \ref{normalization_Lemma}, one can now prove local-in-time existence of smooth solutions to \eqref{Euler3} using Kato's existence theory \cite{Kato}, (see also \cite[chapter 16.2]{Taylor}). Once it is shown that an initially positive $w$ stays positive under evolution by \eqref{Euler3}, the existence of a smooth solution (local in time) of \eqref{Euler1} follows as well.

\section{The Problem of Damping Proportional to Particle-Number Density}
\label{Sec: Euler_particle-number}

In this section we consider a damping term proportional to the particle-number density and show that the  relativistic balance of momentum equations does not reduce to its non-relativistic analog \eqref{Euler_cl_mom}, from which we conclude that such a damping is unphysical. Naively, a damping proportional to particle-number density seems reasonable, since the particle number can be interpreted as rest mass. To begin, consider the energy-momentum tensor of a perfect fluid, \eqref{T_perfect_fluid},  with an equation of state 
\beq \label{eqn_of_state_2}
\epsilon=\epsilon(n,s),
\eeq 
where $\epsilon$ is as before the mass-energy-density, $n$ denotes the \emph{particle-number} density and $s$ denotes the specific entropy density. For the above equation of state, the pressure is given by 
\beq \label{2nd-law}
p = n \frac{\partial \epsilon}{\partial n} - \epsilon .
\eeq
In this framework, we propose the relativistic Euler equation with a \emph{particle-number-damping} as
\begin{eqnarray} 
(n u^\sigma)_{,\sigma} &=& 0, \label{particle-flux-conservation} \\
\text{div}\, T &=& K,   \label{particle-number-damping}
\end{eqnarray}
where 
\beq \label{damping_4-force_particle-number}
K^\mu \equiv  -a \gamma(v)  \left( \begin{array}{c} \frac{1}{c} \vec{v}^2 \cr \vec{v} \end{array} \right) n
\eeq
is the Lorentz force of a damping proportional to $n$. Equation \eqref{particle-flux-conservation} is the conservation of particle-number along flow lines. The particle-number density can be identified with the rest-mass density.

\subsection{Their Newtonian Limit}\label{Sec: Newtonian-limit_2}

As in Section \ref{Sec: Splitting_1}, contraction of $\text{div}(T)=K$ along $u^\mu$ and $\Pi^{\mu\nu}$ gives 
\begin{eqnarray} 
\epsilon_{,\sigma} u^\sigma  + \left( \epsilon + p \right) u^\sigma_{\ ,\sigma} = 0, \label{particle-number-damping_splitting_1_aux} \\
\left(  \epsilon + p \right) u^\mu_{\ ,\sigma} u^\sigma + \Pi^{\mu\nu} p_{,\nu} =  \Pi^{\mu\nu} K_\nu . \label{particle-number-damping_splitting_2}
\end{eqnarray}

We now show that \eqref{particle-number-damping_splitting_1_aux} is equivalent to the relativistic conservation of entropy equation, as a result of \eqref{2nd-law}. For this, use the chain rule to write
\beq \nonumber
\epsilon_{,\sigma} u^\sigma = \frac{\partial \epsilon}{\partial n} n_{,\sigma} u^\sigma + \frac{\partial \epsilon}{\partial s} s_{,\sigma} u^\sigma .
\eeq 
Substituting this and \eqref{2nd-law} into \eqref{particle-number-damping_splitting_1_aux}, we obtain
\beq \nonumber
\frac{\partial \epsilon}{\partial n} \left( n_{,\sigma} u^\sigma + n u^\sigma_{,\sigma} \right) + \frac{\partial \epsilon}{\partial s} s_{,\sigma} u^\sigma = 0,
\eeq
which, by \eqref{particle-flux-conservation}, is equivalent to
\beq \label{particle-number-damping_splitting_1}
s_{,\sigma} \,u^\sigma = 0 .
\eeq
It is now easy to show that the Newtonian limit of \eqref{particle-number-damping_splitting_1} is given by
\beq \label{cons_entropy_non-rel}
\partial_t s + \vec{v} \cdot \nabla s = 0, 
\eeq 
which is the non-relativistic conservation of entropy equation. By the first law of Thermodynamics, \eqref{cons_entropy_non-rel} is equivalent to the non-relativistic conservation of energy equation. Therefore, in the Newtonian limit, $\epsilon$ has the interpretation of energy density and cannot be interpreted as mass density.

We now compute the Newtonian limit ($c \rightarrow \infty$) of \eqref{particle-flux-conservation} and \eqref{particle-number-damping_splitting_2}, beginning with \eqref{particle-flux-conservation}. Observe that
\beq \nonumber
n_{,\sigma} u^\sigma = \frac{\gamma(v)}{c} \left( \partial_t n + \vec{v} \cdot \nabla n  \right),
\eeq
from which together with \eqref{identity_1} we conclude that \eqref{particle-flux-conservation} implies
\beq\nonumber
\gamma(v)\left(  \partial_t n + \vec{v} \cdot \nabla n + n \nabla \cdot \vec{v}  \right) + O(c^{-2}) = 0,
\eeq
which reduces to 
\beq \label{cons_particles_non-rel}
\partial_t n + \nabla \cdot (n\vec{v}) = 0,
\eeq 
as $c\rightarrow \infty$. This coincides with the non-relativistic conservation of mass equation, \eqref{Euler_cl_mass}, which allows us to interpret $n$ as rest mass density.

The above interpretation of $\epsilon$ as energy density and of $n$ as mass density, indicates that the Newtonian limit of \eqref{particle-number-damping_splitting_2} could only agree with the non-relativistic balance of momentum \eqref{Euler_cl_mom}, if $\epsilon$ were proportional to $n$. A dimensional consideration further suggest that the Newtonian limit were only correct, if $\epsilon = c^2 n$ were true. (In fact, assuming $\epsilon = c^2 n$, a straightforward computation shows that \eqref{particle-number-damping_splitting_2} reduces to \eqref{Euler_cl_mom}.) However, since $\epsilon$ is assumed in \eqref{eqn_of_state_2} to be an arbitrary function of $n$ and $s$, we take the above considerations as strong indication that the damping in \eqref{particle-flux-conservation} - \eqref{damping_4-force_particle-number} is \emph{not} physical.

\section{Conclusion}

We introduce a damping term for the relativistic Euler equations in Minkowski spacetime which is proprtional to mass-energy density and we prove that the Newtonian limit of the resulting equations reduce to the correct non-relativistic system. We write the system in symmetric hyperbolic form, (using the Makino variable to replace the mass-energy density), so that in principal Kato's result gives local existence of a smooth solution. We finally prove that the equations with a damping term proportional to particle-number density does not reduce to the correct system in the Newtonian limit, from which we conclude that such a damping seems unphysical.

\section*{Acknowledgements}
I am grateful to Prof. Hermano Frid (IMPA) for suggesting to work on the subject of the Relativistic Damped Euler equations and for introducing me to the important work in \cite{Sideris}.

\begin{appendix}

\section{Some Basics of Special Relativity}\label{Sec: Preliminaries}

We give here a brief summary of the Special Relativity required for this paper, c.f. \cite{Choquet,Weinberg} for a more comprehensive introduction to Relativity.
Special Relativity requires the equations of physics to be Lorentz invariant. For this, the Euclidean space of Newtonian physics must be replaced by the so-called Minkowski spacetime $\R^{1,3}$, which is $\R^4$ endowed with the Minkowski metric
\beq \label{Minkowski-metric}
ds^2 = -c^2 dt^2 + dx^2 +dy^2 +dz^2  = \eta_{\mu\nu} dx^\mu dx^\nu .
\eeq  
Here $(x,y,z)$ denote Cartesian coordinates on $\R^3$, $t$ is the universal time of Newtonian physics and $x^\mu$ (for $\mu \in \{0,1,2,3\}$) denote $(x^0,x^1,x^2,x^3)=(ct,x,y,z)$. In the coordinates $x^\mu$ and in all coordinates related to $x^\mu$ by Lorentz transformations, the non-zero components of $\eta_{\mu\nu}$ are given by the diagonal elements $\eta_{00}=-1$ and $\eta_{11}=1=\eta_{22}=\eta_{33}$.

In Special Relativity the trajectory of a particle in Euclidean space, $t \mapsto \vec{x}(t)   \in \R^3$, 
is replaced by its so-called world line in Minkowski spacetime,    
\beq\nonumber
t \mapsto x^\mu(t) =  \left( \begin{array}{c}  ct \cr \vec{x}(t) \end{array} \right)  \ \ \  \in \R^{1,3}.
\eeq 
It was Einstein's deep insight that the time parameter $t$ has no universal physical meaning, only the so-called \emph{proper time}, which is the Lorentz invariant scalar function defined by
\beq \nonumber 
\tau(t) \, \equiv \, \frac{1}{c}\int^t \sqrt{-\eta(v,v)} dt \, =\, \int^t \sqrt{1-\frac{\vec{v}^2}{c^2}} dt ,
\eeq 
where 
\beq \nn
\vec{v}= \frac{d\vec{x}}{d t} 
\ \ \ \ \ \ \text{and} \ \ \ \ \ \
v^\mu = \frac{d x^\mu}{d t} = \left( \begin{array}{c} c \cr \vec{v}  \end{array} \right)  
\eeq
are the classical and the four-velocity respectively. To clarify, $\tau$ is the time elapsed between two events measured by an observer moving with velocity $\vec{v}$, while $t$ is the time elapsed between the same two events measured by an observer at rest with respect to the coordinates $x^\mu$.    

To obtain a Lorentz invariant velocity of the particle trajectory, we introduce the $4$-velocity as
\beq\nonumber
u^\mu(\tau) \equiv \frac{1}{c}\frac{d x^\mu}{d \tau},
\eeq
which is dimensionless and Lorentz-invariant, since $\tau$ is Lorentz-invariant. The definition of proper time now implies that
\beq\label{proper_time_conversion}
\frac{dt}{d\tau} = \gamma(v), 
\hspace{.5cm} \text{for} \hspace{.5cm} 
\gamma(v) = \left(1- \frac{\vec{v}^2}{c^2}\right)^{-\frac{1}{2}} ,
\eeq 
so that
\beq \label{Prel_relation_u_v}
u^\mu = \frac{\gamma(v)}{c} v^\mu = \frac{\gamma(v)}{c}\left( \begin{array}{c} c \cr \vec{v}  \end{array} \right),
\eeq 
from which we find that the four-velocity is normalized to
\beq\nonumber
u^\sigma u_\sigma \equiv \eta(u,u) = -1.
\eeq
By comparison of the above equations, we find that one can express $\vec{v}$ in terms of $u^\mu$ alone by 
\beq \label{Prel_relation_u_v_3D}
\vec{v}^i = c\frac{u^i}{u^0} ,
\eeq
for $i=1,2,3$. Thus, since $u^\mu$ is Lorentz-invariant and therefore observer independent, we consider $u^\mu$ as the fundamental physical quantity and $\vec{v}$ as being derived from it.

For Newton's $2^{nd}$ law, the equation of motion $\frac{d}{dt} mv = \vec{F}$, (for $\vec{F}$ denoting some force), to be made Lorentz invariant, the time derivative must be replaced by a derivative with respect to proper time $\tau$ and $\vec{v}$ by the $4$-velocity. The Special Relativistic version of Newton's $2^{nd}$ law is thus given by
\beq \label{Newton_eqn_motion_rel}
\frac{d}{d\tau} \left( m_0 u^\nu  \right) = K^\nu,
\eeq
where $m_0$ is a scalar, called the \emph{rest mass} of the particle and $K^\nu$ is the so-called \emph{Lorentz force}. To derive $K^\nu$ in terms of $\vec{F}$, use \eqref{proper_time_conversion} - \eqref{Prel_relation_u_v_3D} to write \eqref{Newton_eqn_motion_rel} as 
\beq \nonumber
 \gamma(v) \frac{d}{dt} \left( m_0 \frac{\gamma(v)}{c}  \vec{v}^i  \right) = K^i,
\eeq
for $i=1,2,3$. The above equation agrees with Newton's $2^{nd}$ law of motion provided
\begin{eqnarray}
m &=&  \frac{\gamma(v)}{c} m_0 ,  \nonumber \\ 
K^i &=& \gamma(v) \vec{F}^i, \ \ \ \ \ \ \text{for} \ i=1,2,3.   \label{4-force_spatial}
\end{eqnarray}
Moreover, assuming that $m_0$ is constant, contracting \eqref{Newton_eqn_motion_rel} with $u_\nu$ and using that $u^\nu u_\nu = -1$, we find that the resulting expression on the left hand side vanishes, so that \eqref{4-force_spatial} for the expression on the right hand side finally yields   
\beq \label{4-force_time}
K^0 = \frac{\gamma(v)}{c} \vec{F}\cdot \vec{v}.
\eeq

\end{appendix}

\providecommand{\bysame}{\leavevmode\hbox to3em{\hrulefill}\thinspace}
\providecommand{\MR}{\relax\ifhmode\unskip\space\fi MR }
\providecommand{\MRhref}[2]{%
  \href{http://www.ams.org/mathscinet-getitem?mr=#1}{#2} }
\providecommand{\href}[2]{#2}

\end{document}